\documentclass[conference]{IEEEtran}
\IEEEoverridecommandlockouts
\usepackage{cite}
\usepackage{amsmath,amssymb,amsfonts}
\usepackage{algorithmic}
\usepackage{graphicx}
\usepackage{textcomp}
\usepackage{xcolor}
\usepackage{booktabs}
\usepackage{multirow} 
\def\BibTeX{{\rm B\kern-.05em{\sc i\kern-.025em b}\kern-.08em
    T\kern-.1667em\lower.7ex\hbox{E}\kern-.125emX}}
\begin{document}

\title{EENED: End-to-End Neural Epilepsy Detection based on Convolutional Transformer\\

\thanks{{$^{*}$Corresponding Author: Yang Liu and Xinliang Zhou.}
\\
{Chenyu Liu, and Xinliang Zhou are with the School of Computer Science and Engineering, Nanyang Technological University, 639798, Singapore {\tt\small\{chenyu003, xinliang001\}@e.ntu.edu.sg} }%

{$^{2,1}$Yang Liu is with Zhejiang Sci-Tech University, Hangzhou, 314423, China and the School of Computer Science and Engineering, 50 Nanyang Avenue, 639798, Singapore
        {\tt\small \ {yangliu}@ntu.edu.sg}}%
}}

 \author{Chenyu Liu$^1$\hspace{1cm}$^{*}$Xinliang Zhou$^1$\hspace{1cm}$^{*}$Yang Liu$^2{}^,{}^1$\\

$^1$School of Computer Science and Engineering, Nanyang Technological University\\
$^2$Zhejiang Sci-Tech University\\

\{chenyu003, xinliang001\}@e.ntu.edu.sg, 
\{yangliu\}@ntu.edu.sg
}

\maketitle
\begin{abstract}
Recently Transformer and Convolution neural network (CNN) based models have shown promising results in EEG signal processing. Transformer models can capture the global dependencies in EEG signals through a self-attention mechanism, while CNN models can capture local features such as sawtooth waves.
In this work, we propose an end-to-end neural epilepsy detection model, EENED, that combines CNN and Transformer. Specifically, by introducing the convolution module into the Transformer encoder, EENED can learn the time-dependent relationship of the patient's EEG signal features and notice local EEG abnormal mutations closely related to epilepsy, such as the appearance of spikes and the sprinkling of sharp and slow waves.  Our proposed framework combines the ability of Transformer and CNN to capture different scale features of EEG signals and holds promise for improving the accuracy and reliability of epilepsy detection. Our source code will be released soon on GitHub.
\end{abstract}


\section{Introduction}

Epilepsy is a chronic non-infectious disease caused by the paroxysmal abnormal hypersynchronous electrical activity of brain neurons that affects people of all ages\cite{casson2010wearable}. It is also one of the most common neurological diseases in the world. Due to brain abnormalities, there are differences in the starting position and transmission mode of electrical activity, and the clinical manifestations of epilepsy are characterized by diversification and complexity. Repeated epileptic seizures will cause persistent adverse effects on patients' mental and cognitive functions and even endanger their lives. Studies have shown that EEG signals in epileptic patients differ from non-epileptic subjects. Therefore, judging whether a subject suffers from epilepsy by identifying EEG signals has important clinical significance for diagnosing epilepsy\cite{zhou2023interpretable}.

Neural network models based on Transformer and CNN are widely used in epilepsy detection tasks via EEG. Transformer performs very well in fields such as natural language processing (NLP), and it can capture long-distance dependencies in temporal signals through self-attention mechanisms. Therefore, Transformer is also widely used in EEG signal processing, modeling EEG signals through the self-attention mechanism, and extracting spatio-temporal relationship features related to epilepsy in the signal. In contrast, CNN-based neural networks pay more attention to capturing local features when processing time-series data, such as signal waveforms of different frequencies of EEG signals. Through hierarchical convolution operations, the convolutional network extracts high-dimensional representations of EEG signals for epilepsy detection and classification. In summary, the neural networks based on Transformer and CNN have different advantages in processing time series signals and can extract features of different scales in epilepsy detection tasks.

\begin{figure}
\centering 
\includegraphics[width=0.70 \columnwidth]{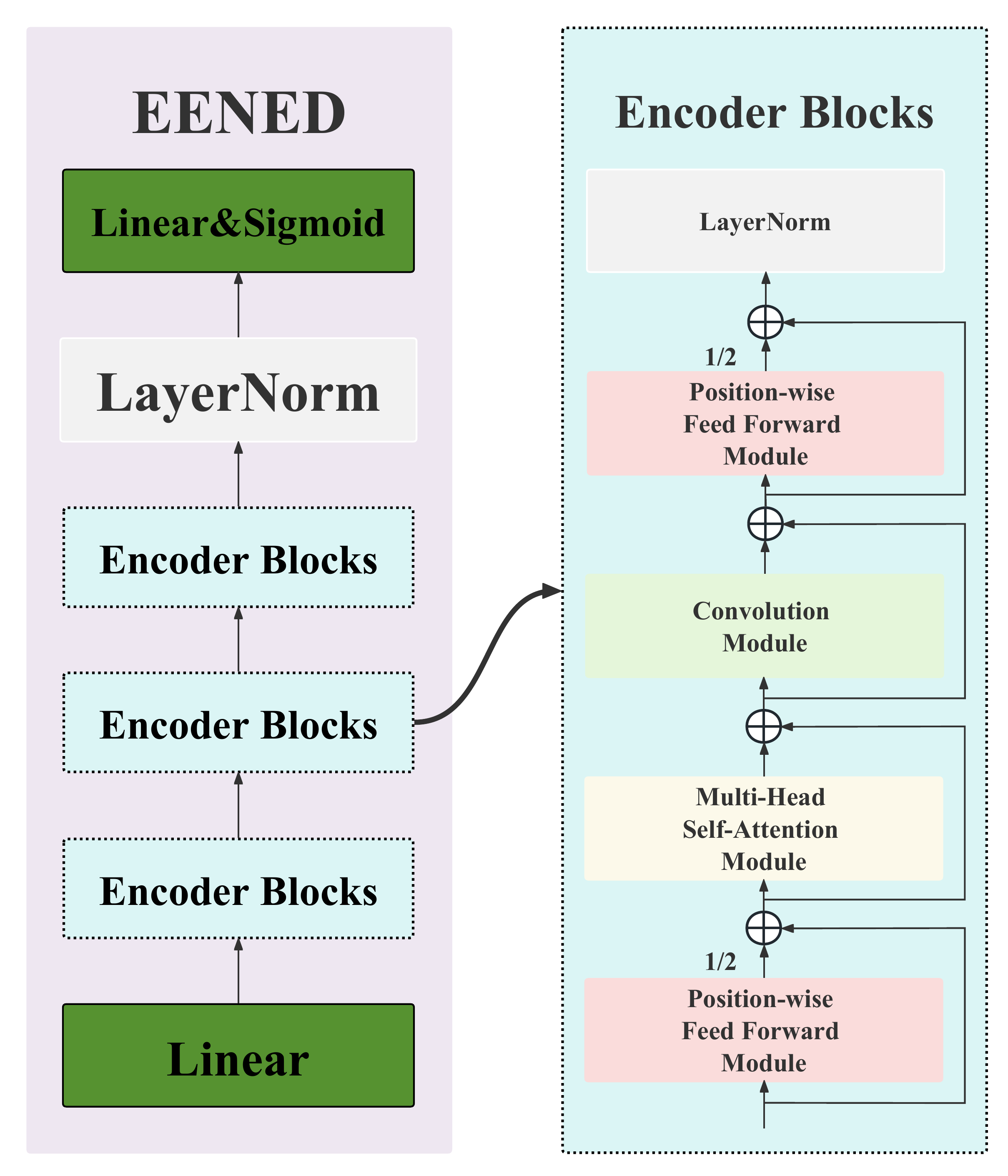}
\caption{\textbf{End-to-End Neural Epilepsy Detection model architecture.} EENED's structure contains several encoder blocks and linear layers. The encoder's multi-headed self-attention and convolution modules are sandwiched between two macaron-like positional-wise feed-forward layers with half-step residual connection} 
\label{fig:eened} 
\end{figure}

Since global and local interactions are all essential for parameter efficiency, we propose an End-to-End neural network based on the convolutional transformer encoder structure, as shown on the left side of Fig \ref {fig:eened}. Inspired by \cite{gulati2020conformer}, the encoder blocks of EENED follow the self-attention mechanism of the Transformer model to extract the time-dependent relationship in time-series EEG signals. Unlike the traditional Transformer, we removed the positional encoding \cite{fujita2019end} in the multi-head attention mechanism and divided the feed-forward layer into two parts to form a sandwich structure. At the same time, a convolution module is introduced in the encoder blocks to enhance the extraction of local features. This convolution module is a hybrid convolution module, including one-dimensional depth-wise convolution and one-dimensional point-wise convolution, which is used to analyze local features in EEG signals that may be related to epilepsy, such as abnormal waveforms and sudden changes in frequency and amplitude.

As proved by the experimental results of the Epileptic Seizure Recognition dataset\cite{andrzejak2001indications}, EENED shows higher accuracy than CNN and Transformer-based neural networks. While capturing the long-term dependence of temporal EEG signals and extracting local signal features through convolution modules, EENED has shown reliability and great potential in epilepsy detection tasks.                                  

\section{Methodology}

\subsection{Encoder Blocks}

Inspired by Macaron-Net \cite{lu2019understanding}, the encoder blocks of EENED adopt a sandwich structure, which divides the feedforward layer in the Transformer encoder into two half-step residual feedforward units. A multi-head self-attention and convolution modules are included between the two feedforward modules, as shown in Fig \ref{fig:eened}. Mathematically, for given input $(f_{(e-1)}^{t}|t=1,...,T)$ to the $e^{th}$ encoder block, the output $(f_{(e)}^{t}|t=1,...,T)$ of the block is:

{\footnotesize
\begin{equation}
\begin{aligned}
f^{(e)}_{FF_1}= (f_{(e-1)}^{t}|t=1,...,T) + \frac{1}{2}\operatorname{PWFF}((f_{(e-1)}^{t}|t=1,...,T))
\end{aligned}
\end{equation}
\begin{equation}
\begin{aligned}
f^{(e)}_{MHSA}=f^{(e)}_{PWFF_1} + \operatorname{MHSA}(f^{(e)}_{PWFF_1})
\end{aligned}
\end{equation}
\begin{equation}
\begin{aligned}
f^{(e)}_{Conv}=f^{(e)}_{MHSA} + \operatorname{Conv}(f^{(e)}_{MHSA})
\end{aligned}
\end{equation}
\begin{equation}
\begin{aligned}
(f_{(e)}^{t}|t=1,...,T)=\operatorname{LayerNorm}(f^{(e)}_{Conv}+\frac{1}{2}\operatorname{PWFF}(f^{(e)}_{Conv}))
\end{aligned}
\end{equation}
}

Where $\operatorname{PWFF}()$ refers to the position-wise feed-forward module, $\operatorname{MHSA}()$ refers to the multi-head self-attention module, and $\operatorname{Conv}()$ refers to the convolution module.

\subsubsection{Position-wise feed-forward module(PWFF)}

\begin{figure}
\centering 
\includegraphics[width=1.00\columnwidth]{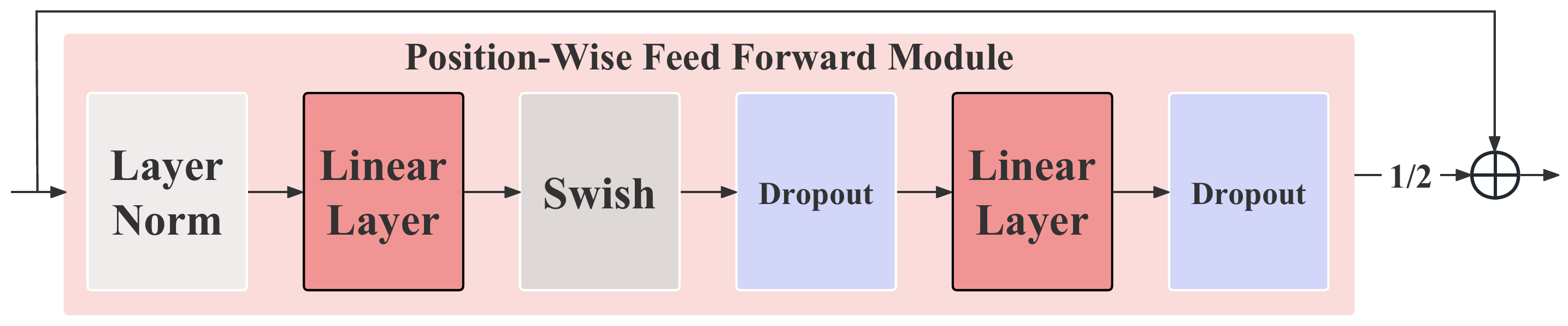}
\caption{\textbf{Position-wise feed-forward model.} Two linear layers increase and decrease the dimensionality of the data, respectively.} 
\label{fig:feedforward} 
\end{figure}

The Transformer architecture\cite{vaswani2017attention} employs a feed-forward module before and after the MHSA module, consisting of two linear layers and activation as shown in Fig \ref{fig:feedforward}. The $e^{th}$ PWFF module transforms a sequence of input vectors $(f_{(e-1)}^{t}|t=1,...,T)$ from the previous encoder as follows:

{\footnotesize
\begin{equation}
\begin{aligned}
F^{(e)}= \operatorname{LayerNorm}([f_{t}^{(e-1)} \cdots f_{T}^{(e-1)}])\: \epsilon \:\mathbb{R}^{T\times D}
\end{aligned}
\end{equation}
\begin{equation}
\begin{aligned}
\bar{F}^{(e)}= \operatorname{Swish}(F^{(e)}W_{1}^{(e)}+1b_{1}^{(e)^{\top }})W_{2}^{(e)}+1b_{2}^{(e)^{\top }}\: \epsilon \:\mathbb{R}^{T\times D}
\end{aligned}
\end{equation}
}

Where $T$ is the length of time and $D$ is the feature dimension. $W_{1}^{(e)}\: \epsilon \:\mathbb{R}^{D\times d_{pwff}}$ and $b^{(e)}_{1}\: \epsilon \:\mathbb{R}^{d_{pwff}}$ are the projection matrix and bias of the first linear layer, $1\: \epsilon \:\mathbb{R}^{T} $ is an all-one vector. $d_{pwff}$ is the dimension of hidden units. We also apply Swish activation \cite{ramachandran2017searching} $\operatorname{Swish}(\cdot)$ and dropout\cite{srivastava2014dropout} to help regularize the module. $W_{2}^{(e)}\: \epsilon \:\mathbb{R}^{D\times d_{pwff}}$ and $b^{(e)}_{2}\: \epsilon \:\mathbb{R}^{d_{pwff}}$ are the second linear projection matrix and bias.
Following the pre-norm residual units\cite{wang2019learning}, the final output of the PWFF module is computed as follows:

{\footnotesize
\begin{equation}
\begin{aligned}
{F}^{(e)}_{PWFF} = F^{(e)}+ \frac{\operatorname{Dropout}(\bar{F}^{(e)})}{2}\: \epsilon \:\mathbb{R}^{T\times D}
\end{aligned}
\end{equation}
}

\subsubsection{Multi-head self-attention module(MHSA)}

\begin{figure}
\centering 
\includegraphics[width=0.80 \columnwidth]{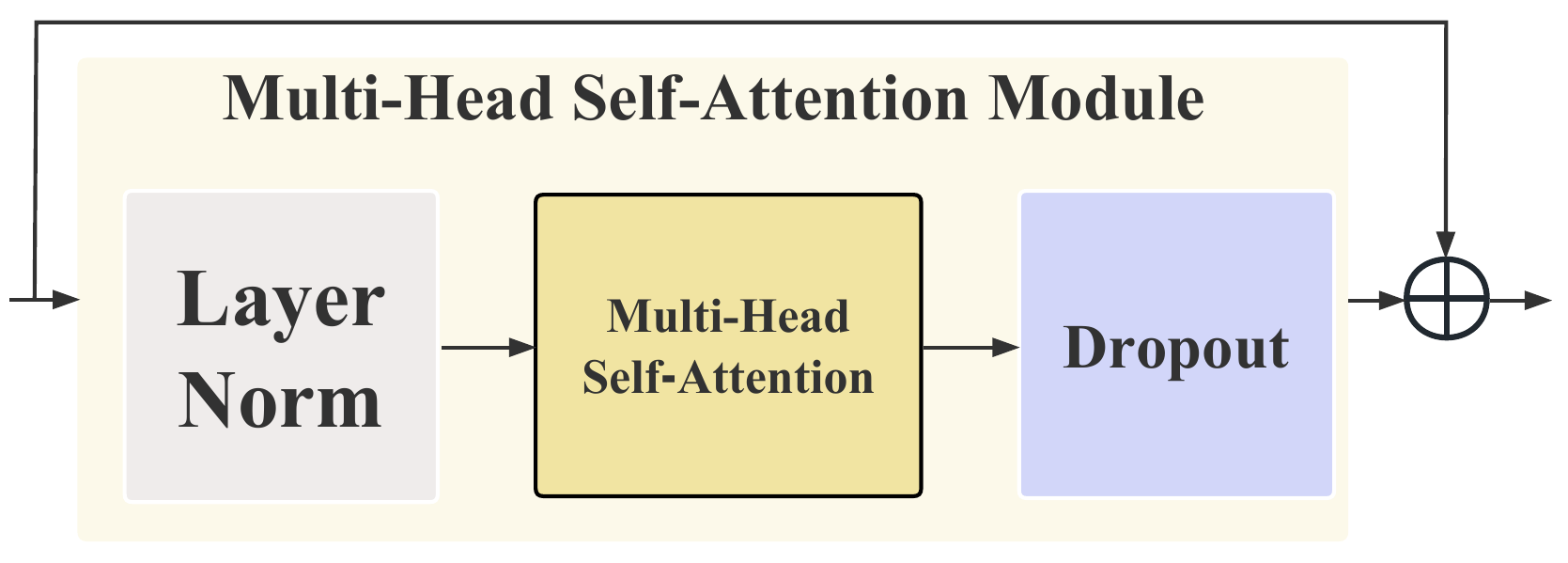}
\caption{\textbf{Multi-head self-attention module.} We use a multi-head self-attention similar to the transformer encoder but remove the relative positional embedding in this pre-norm residual unit.} 
\label{fig:mhsa} 
\end{figure}

The structure of the multi-head self-attention module is shown in Fig \ref{fig:mhsa}.
The MHSA module in the $e^{th}$ encoder processes features from the PWFF module. The input features ${F}^{(e)}_{PWFF}$ is converted by layer normalization:

{\footnotesize
\begin{equation}
\begin{aligned}
\bar{F}^{(e)}= \operatorname{LayerNorm}({F}^{(e)}_{PWFF})\: \epsilon \:\mathbb{R}^{T\times D}
\end{aligned}
\end{equation}
}

Then, in the multi-head self-attention block, each attention head computes a pairwise similarity matrix $S_{h}^{(e)}$ using the dot products of query vectors $\bar{F}^{(e)}Q_{h}^{(e)}\: \epsilon \:\mathbb{R}^{T\times d}$ and key vectors $\bar{F}^{(e)}K_{h}^{(e)}\: \epsilon \:\mathbb{R}^{T\times d}$

{\footnotesize
\begin{equation}
\begin{aligned}
S_{h}^{(e)}=\bar{F}^{(e)}Q_{h}^{(e)}(\bar{F}^{(e)}K_{h}^{(e)})^{\top }\: \epsilon \:\mathbb{R}^{T\times T}(1\leq h \leq H)
\end{aligned}
\end{equation}
}

where $H$ is the number of heads. $Q_{h}^{(e)}$,$K_{h}^{(e)}\: \epsilon \:\mathbb{R}^{D\times d}$ are query and key projection matrices for the $h^{th}$ head. The pairwise similarity matrix $S_{h}^{(e)}$ is scaled by $1/\sqrt{D/H}$ and a softmax function is applied to form the attention weight matrix $A_{h}^{(e)}$:

{\footnotesize
\begin{equation}
\begin{aligned}
A_{h}^{(e)}=\operatorname{Softmax}\left ( \frac{S_{h}^{(e)}}{\sqrt{D/H}} \right )\: \epsilon \:\mathbb{R}^{T\times T}
\end{aligned}
\end{equation}
}

Then the attention weight matrix $A_{h}^{(e)}$ is used to compute context vectors $C_{h}^{(e)}$ with the value vectors $\bar{F}^{(e)}V_{h}^{(e)}\: \epsilon \:\mathbb{R}^{T\times d}$:

{\footnotesize
\begin{equation}
\begin{aligned}
C_{h}^{(e)}=A_{h}^{(e)}(\bar{F}^{(e)}V_{h}^{(e)})\: \epsilon \:\mathbb{R}^{T\times d}
\end{aligned}
\end{equation}
}

where $V_{h}^{(e)}\: \epsilon \:\mathbb{R}^{D\times d}$ is the value projection matrix. The final output feature of the multi-head self-attention module is computed by the concatenation of all heads' context vectors and an output projection matrix $O^{(e)}\: \epsilon \:\mathbb{R}^{D\times D}$:

{\footnotesize
\begin{equation}
\begin{aligned}
\bar{F}^{(e)}_{MHSA}=[C_{1}^{(e)}\cdots C_{H}^{(e)}]O^{(e)}\: \epsilon \:\mathbb{R}^{T\times D}
\end{aligned}
\end{equation}
\begin{equation}
\begin{aligned}
{F}^{(e)}_{MHSA}= \operatorname{LayerNorm}(\bar{F}^{(e)} + \operatorname{DropOut}(\bar{F}^{(e)}_{MHSA}))\: \epsilon \:\mathbb{R}^{T\times D}
\end{aligned}
\end{equation}
}

\subsubsection{Convolution module}

\begin{figure*} 
\centering 
\includegraphics[width=1.7 \columnwidth]{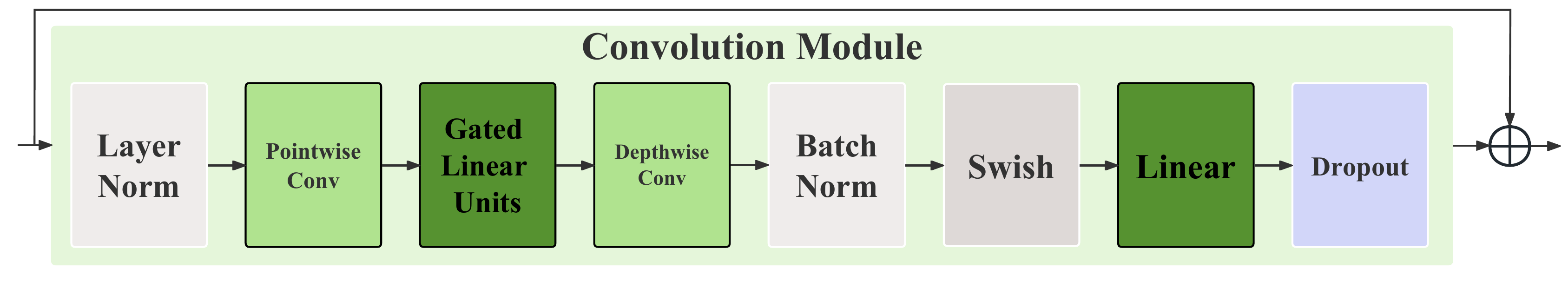}
\caption{\textbf{Convolution module.} The convolution module contains two convolutional layers of different scales, with Gated linear units used as the activation layer in the middle. The Swish activation function is used, followed by a linear layer.} 
\label{fig:conv} 
\end{figure*}

Similar to \cite{wu2020lite}, the convolution module, as shown in Fig \ref{fig:conv}, takes $F^{(e)}_{MHSA}$ as input and starts with a  point-wise convolution and a gated linear unit (GLU)\cite{dauphin2017language}:

{\footnotesize
\begin{equation}
\begin{aligned}
\bar{F}^{(e)}= \operatorname{PWConv}(\operatorname{LayerNorm}({F}^{(e)}_{MHSA}))\: \epsilon \:\mathbb{R}^{T\times 2D}
\end{aligned}
\end{equation}
\begin{equation}
\begin{aligned}
\bar{F}^{(e)}_{glu}= (\hat{F}^{(e)}W_{1}^{(e)}+b_{1}^{(e)})\bigotimes \sigma (\check{F}^{(e)}W_{2}^{(e)}+b_{2}^{(e)})\: \epsilon \:\mathbb{R}^{T\times D}
\end{aligned}
\end{equation}
}

where $\operatorname{PWConv}(\cdot)$ is a one-dimensional point-wise convolutional layer with kernel size (1 X 1) and stride of 1. The output $\bar{F}^{(e)}\: \epsilon \:\mathbb{R}^{T\times 2D}$ could be divided into $\hat{F}^{(e)}\: \epsilon \:\mathbb{R}^{T\times D}$ and $\check{F}^{(e)}\: \epsilon \:\mathbb{R}^{T\times D}$, which are the first half and the second half of the output, respectively. $W_{1}^{(e)}\: \epsilon \:\mathbb{R}^{D\times D}$, $b_{1}^{(e)}\: \epsilon \:\mathbb{R}^{D}$,$W_{2}^{(e)}\: \epsilon \:\mathbb{R}^{D\times D}$, $b_{1}^{(e)}\: \epsilon \:\mathbb{R}^{D}$ are learned parameters, $\sigma$ is the sigmoid function and $\bigotimes$ is the element-wise product. The output of GLU $\bar{F}^{(e)}_{glu}$ is processed with a $\operatorname{DWConv}(\cdot)$:

{
\begin{equation}
\begin{aligned}
\bar{F}^{(e)}_{DWConv}= W_{conv}^{(e)}\operatorname{Swish}(\operatorname{DWConv}(\bar{F}^{(e)}_{glu}))+b_{Conv}^{(e)}\: \epsilon \:\mathbb{R}^{T\times D}
\end{aligned}
\end{equation}
}

where $W_{Conv}^{(e)}\: \epsilon \:\mathbb{R}^{d_{pwconv}\times D}$, $b_{Conv}^{(e)}\: \epsilon \:\mathbb{R}^{D}$ are learned linear parameters of Convolution module. $\operatorname{Swish}(\cdot)$ is the activation function. $\operatorname{DWConv}(\cdot)$ is a one-dimensional depth-wise convolutional layer with a kernel size of 15. $d_{pwconv}$ is the dimension of depth-wise convolution layer output. The final output features ${F}^{(e)}_{Conv}$ of the convolution module are as follows:

{
\begin{equation}
\begin{aligned}
{F}^{(e)}_{Conv} = {F}^{(e)}_{MHSA}+ \operatorname{Dropout}(\bar{F}^{(e)}_{DWConv})\: \epsilon \:\mathbb{R}^{T\times D}
\end{aligned}
\end{equation}
}

\section{Experiments and Results}
\begin{figure*}
\centering 
\includegraphics[width=1.80\columnwidth]{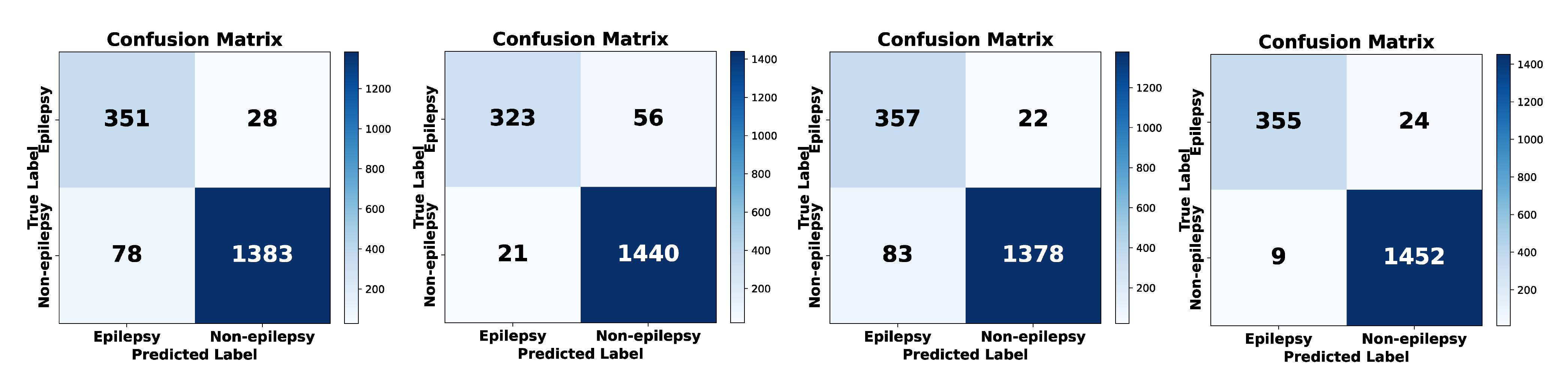}
\caption{Confusion matrix of the predicted results of the four models. From left to right: Dense-CNN, CNN-LSTM, Transformer, and EEENED.} 
\label{fig:result} 
\end{figure*}

\subsection{Dataset}
The Epileptic Seizure Recognition dataset\cite{andrzejak2001indications} contains EEG recordings from 500 subjects. The brain activity of each subject was recorded for 23.6 seconds. Since four of the five categories are unrelated to epilepsy, we reduced the labels to one category. The training set contains 7360 segments of EEG signal data, and the test set contains 1840 segments of EEG signal data, of which 1461 are non-epileptic EEG signals. We followed the data processing in \cite{eldele2021time}.

\subsection{Model configuration}

\subsubsection{EENED}
EENED contains three Encoder blocks, and each Encoder layer contains a self-attention module and a convolution module. The attention mechanism consists of 8 attention heads, each with a dimension of 64. The convolution module uses one-dimensional convolution layers and GLU activation function; the convolution kernel size is 15, the step size is 1, the padding is 7, and the number of groups is 512. The model also contains two feed-forward layers and residual connections, and finally, it applies two linear layers and a sigmoid activation function to classify the features.

\subsubsection{Dense-CNN}
Dense-CNN\cite{saab2020weak} is comprised of multiple convolutional layers, with each layer having a different set of convolutional filters. The first is a linear layer that outputs a tensor of size 512. The subsequent layers use an architecture called Dense-Inception which is made up of several Inception modules. Each Inception module consists of multiple branches that process the input in parallel using different convolutional filters of different kernel sizes. The output of each branch is then concatenated along the channel dimension and fed into a 1x1 convolutional layer to reduce the number of channels. The output of one Inception module is then fed into the next Inception module, and the process is repeated until the final layer, which outputs a tensor of size 18.

\subsubsection{Transformer}
This network architecture is a Transformer-based classifier consisting of an upsample layer, two linear layers, Transformer-Encoder layers with 3 Transformer encoders, a fully connected layer, and a sigmoid activation function. The upsample layer maps the input sequence to a hidden state of size 512. The first linear layer maps the hidden state to a single value, which is then used to scale the output of the Transformer-Encoder layer. The Transformer-Encoder layer contains a self-attention mechanism and two linear and dropout layers. The self-attention mechanism uses 8 attention heads and an output projection matrix of size 512. The dropout probability is set to 0.1 for the attention and linear layers. The fully connected layer maps the output of the Transformer-Encoder to a single value, which is then passed through the sigmoid activation function to produce the final classification output. 

\subsubsection{CNN-LSTM}
CNN-LSTM\cite{ahmedt2020neural} comprises a convolutional neural network and a long and short-term memory (LSTM) layer. The convolutional neural network consists of two convolutional layers and a maximum pooling layer, and the output size of the fully connected layer is 512. The input of the LSTM layer is the output of the fully connected layer and contains two LSTM layers with a hidden state size of 128.

\subsection{Result and Analysis}

As shown in Fig \ref{fig:result}, from the prediction result confusion matrix of the four models, the accuracy rates of Dense-CNN and Transformer are 0.942 and 0.943, respectively. The accuracy rates of CNN-LSTM and EENED are 0.958 and 0.982, respectively. Among them, EENED achieved the highest accuracy rate. When Dense-CNN is used to process time-domain signals, it cannot perform better due to the lack of a mechanism for processing sequence data. Transformer is good at global modeling of time-domain data. However, it is challenging to capture short sequence signal features related to epilepsy in EEG signals, such as the appearance of spikes and the sprinkling of sharp and slow waves.
In contrast, CNN-LSTM integrates local and global feature extraction to a certain extent, which can effectively capture the temporal dependence of time series data. EENED has the highest accuracy rate in the experimental results. It combines the characteristics of CNN and Transformer. While learning the global temporal dependence of epilepsy-related EEG signals, it captures local signal features through the convolution module.

\begin{table}
\centering
\caption{Comparison of performance (F1 Score and Accuracy) of four neural networks on The Epileptic Seizure Recognition dataset. }
\begin{tabular}{lll}
\hline
\textbf{Network} & \textbf{Accuracy} & \textbf{F1 Score} \\ \hline
Dense-CNN\cite{saab2020weak}        & 0.942             & 0.963             \\
Transformer      & 0.943             & 0.963             \\
CNN-LSTM\cite{ahmedt2020neural}         & 0.958             & 0.974             \\
EENED            & \textbf{0.982}    & \textbf{0.989}    \\ \hline
\end{tabular}
\end{table}

\section{Conclusion}

EENED achieves higher accuracy in epilepsy detection tasks than other transformer and CNN-based neural networks. The experimental results show that EENED can combine the ability of the self-attention mechanism to build the long-term dependence of EEG signals and the characteristics of the convolution module to extract local EEG signal features, which is crucial for using EEG to detect epilepsy as a reference for medical diagnosis.

\section{ACKNOWLEDGMENTS}
This research is supported by National Research Foundation (NRF) Singapore, NRF Investigatorship NRF-NRFI06-2020-0001.

\bibliographystyle{IEEEbib}
\bibliography{ref}

\end{document}